\documentclass[pra, twocolumn,notitlepage,groupedaddress,nofootinbib]{revtex4-1}
\usepackage{braket}
\usepackage[dvipsnames]{xcolor} %
\usepackage{amsmath, amssymb, graphicx, bm}
\usepackage{comment}

\newcommand{\br}{\bm{r}}

\newcommand{\ha}{\hat{a}}
\newcommand{\had}{\hat{a}^\dagger}

\newcommand{\hbr}{\hat{\bm{r}}}
\newcommand{\om}{\bm{0}_M}
\DeclareMathOperator{\tr}{tr}

\DeclareMathOperator{\sech}{sech}
\DeclareMathOperator{\sinc}{sinc}
\DeclareMathOperator{\sign}{sign}
\newcommand{\W}{\hat{\mathcal{W}}}
\newcommand{\HH}{\hat{\mathcal{H}}}
\newcommand{\TT}{\hat{\mathcal{U}}_2}
\newcommand{\Stwo}{\bm{S}_2}

\newcommand{\uu}{\hat{\mathcal{U}}}
\newcommand{\amp}{c}
\usepackage{color}

\usepackage{dsfont}
\usepackage{orcidlink}
\usepackage[normalem]{ulem}

\newcommand{\vp}{\bm{V}_Q}
\newcommand{\vn}{\bm{V}_C}
\newcommand{\pr}[1]{| #1 \rangle \langle #1|}

\begin{document}
	\title{What's my phase again? \\Computing the vacuum-to-vacuum amplitude of quadratic bosonic evolution}
	\author{Nicol\'as Quesada$^{\orcidlink{0000-0002-0175-1688}}$,}
	\email{nicolas.quesada@polymtl.ca}
	\affiliation{%
		D\'epartement de g\'enie physique, \'Ecole polytechnique de Montr\'eal, Montr\'eal, QC, H3T 1J4, Canada
	}%
	\date{\today}
	
	\begin{abstract}
		Quadratic bosonic Hamiltonians and their associated unitary transformations form a fundamental class of operations in quantum optics, modelling key processes such as squeezing, displacement, and beam-splitting. Their Heisenberg-picture dynamics simplifies to linear (or possibly affine) transformations on quadrature operators, enabling efficient analysis and decomposition into optical gate sets using matrix operations. However, this formalism discards a phase, which, while often neglected, is essential for a complete unitary characterization.  
		We present efficient methods to recover this phase directly from the vacuum-to-vacuum amplitude of the unitary, using calculations that scale polynomially with the number of modes and avoid Fock space manipulations. We reduce the general problem for time-dependent Hamiltonians to integration, and provide analytical results for key cases including time-independent Hamiltonians which are positive definite, passive, active, or single-mode. Finally, we show that our results can be easily used to obtain the phase associated with any Gaussian state, be it mixed or pure.
	\end{abstract}
	\maketitle
	
	\section{Introduction}
	
	Quadratic bosonic Hamiltonians and their associated Gaussian unitaries are among the most widely studied operations in quantum optics~\cite{serafini2023quantum}. They describe fundamental processes such as mode squeezing, displacement, beam splitting as well as phase and controlled-phase operations in the continuous-variable model of quantum computation~\cite{lloyd1999quantum} and certain discrete-variable quantum error correction encodings~\cite{gottesman2001encoding}. These transformations can be efficiently analyzed in the Heisenberg picture, where their action on the quadrature operators reduces to linear (or possibly affine) transformations implemented via symplectic matrices (and possibly displacement vectors).
	
	This symplectic formalism enables the decomposition of such transformations into optical gate sets—e.g., beam splitters, squeezers, and displacements—through purely finite-dimensional matrix operations~\cite{braunstein2005squeezing,Reck1994decomposition,Clements2016decomposition,de2018simple,kumar2021unitary,zhou2024bosehedral,LpezPastor2021,bell2021further,arends2024decomposing,yasir2025compactifying,pereira2020minimum,saygin2020robust,idel2015sinkhorn,girouard2025near}. However, a key limitation of this approach is the loss of phase information: two unitaries that act identically on quadratures may still differ by a physically meaningful phase. One can use  Schur’s lemma (cf. Appendices A and B  of Ballentine~\cite{ballentine2014quantum}) to formalize this observation by stating that such unitaries differ by at most a phase when their action is indistinguishable on an irreducible set of operators, such as the bosonic quadratures.
	
	This phase, although often neglected, plays an essential role in several contexts: it appears in superpositions of Gaussian states~\cite{yao2024riemannian,marshall2023simulation,bourassa2021fast,tara1993production} and in controlled-unitary operations relevant to relativistic quantum information~\cite{sachs2017entanglement,funai2025gaussian}, it is also important for the development of quantum information technologies~\cite{didier2015fast,zhu2013circuit} and in topological physics~\cite{pancharatnam1956generalized,wanjura2023quadrature,mukunda1993quantum,anandan1988geometric}. Furthermore, accurate phase accounting is necessary when Gaussian unitaries are combined into larger computations or used as components in approximating non-Gaussian states and processes~\cite{yao2024riemannian,motamedi2025stellar,dias2024classical}. 
	
	We should mention that related results have been very recently derived by Guaita et al.~\cite{guaita2024representation} using tools from representation theory for time-independent (fermionic and bosonic) Hamiltonians. Similarly, Funai~\cite{funai2025gaussian} has addressed related problems in the context of quantum fields for a set of restricted Hamiltonians in quantum field theory. Finally, de Gosson and Nicacio have analyzed this problem in terms of metaplectic isotopies~\cite{de2018relative}.
	Our results address the problem using linear algebra tools that emphasize the use of the Bargmann representation~\cite{bargmann1961hilbert,bargmann1967hilbert,bargmann1971completeness,yao2024riemannian,motamedi2025stellar,xu2025bargmann} (reviewed in Appendices~\ref{app:bargmann} and \ref{app:gaussian}). This representation, unlike the usual one in terms of symplectic matrices, is capable of keeping track of the phase of the composition of two Gaussian unitaries~\cite{yao2024riemannian}. Moreover, our results also allow us to track phases for continuous evolution dictated by time-dependent Hamiltonians.

	In this work, we present efficient methods to recover the phase of quadratic unitaries by computing their vacuum-to-vacuum amplitude. These methods scale polynomially with the number of modes and entirely avoid manipulations in the infinite-dimensional Fock space. We provide analytical results for special cases
	---when the Hamiltonian is positive definite, passive, active, or single-mode--- and 
	show how the general multimode problem can be reduced to integration, even for time-dependent Hamiltonians.

	The manuscript is structured as follows. Sec.~\ref{sec:problem} formalizes the problem and sets up the notation. In Sec.~\ref{sec:Hpos} we present analytical methods for Hamiltonian matrices that can be symplectically diagonalized.
	The general case, including time-dependent Hamiltonians, is addressed in Sec.~\ref{sec:general}, where we show how the vacuum amplitude can be obtained from a one-dimensional integral of a quantity derived from the solution of a set of linear differential equations. 
	Concluding remarks, follow in Sec.~\ref{sec:final}.

	\section{Notation and  Problem statement}\label{sec:problem}
	We setup some basic notation before stating the problem. The vector of quadrature operators of an $M$-mode system is~\cite{serafini2023quantum}
	\begin{align}
		\bm{\hat r} &= 
		\hat{\bm{q}}  \oplus 
		\hat{\bm{p}}
		= (\hat q_1,\ldots,\hat q_M, \hat p_1,\ldots,\hat p_M)^T,
	\end{align}
	where $\hat{\bm{q}}$ is the vector of position operators and $\hat{\bm{p}}$ is the vector of momentum operators. 
	These operators satisfy the canonical commutation relations $[\hat{r}_j,\hat{r}_k] = i \hbar \Omega_{jk}$ where $	
	\bm{\Omega} = \left(\begin{smallmatrix}
		0 & \ \mathbb{I}_M \\
		-\mathbb{I}_M & \ 0
	\end{smallmatrix}\right),$
	is the so-called symplectic matrix, and $\mathbb{I}_M$ is the $M$-dimensional identity matrix.
	It is also useful to introduce annihilation and creation (collectively ladder) operators for mode $j$
	\begin{align}
		\ha_j = \frac{1}{\sqrt{2 \hbar}}(\hat{q}_j+ i \hat{p}_j), \quad \had_j =  \frac{1}{\sqrt{2 \hbar}}(\hat{q}_j- i \hat{p}_j).
	\end{align}
	Finally, we denote the multimode vacuum state by $\ket{\bm{0}} = \ket{0_1} \otimes \ldots \otimes \ket{0_M}$; this is the unique state that satisfies $\ha_j \ket{\bm{0}} = 0 \ \forall j$. 
	
	Note that we denote finite-dimensional vectors by boldface lower case letters, matrices by boldface uppercase letters and infinite-dimensional operators in Fock space with a hat.
	
	The question that concerns this manuscript is motivated by the following problem: given a time-independent quadratic Hamiltonian in quadrature operators~\cite{serafini2023quantum}, specified by an even-dimensional symmetric matrix $\bm{H}$ \footnote{The case where the Hamiltonian has linear terms in the quadrature operators is analyzed in Appendix \ref{app:linear}}, 
	\begin{align}
		\HH = \tfrac{1}{2\hbar} \hbr^T \bm{H} \hbr = \tfrac{1}{2\hbar} \sum_{k=1}^{2M} \sum_{l = 1}^{2M} H_{kl} \hat{r}_k \hat{r}_l ,
	\end{align}
	how can one factor its associated unitary operator 
	\begin{align}\label{eq:unit}
		\uu = \exp(- i\hat{\mathcal{H}}t)=\exp\left(- \tfrac{i}{2 \hbar} \hbr^T \bm{H}  \hbr t \right),
	\end{align}
	into simple one- and two-mode gates, such as squeezing and beamsplitter operations?
	
	Significant progress on this problem can be obtained by looking at the action of the operator in Eq.~\eqref{eq:unit} in the Heisenberg picture. Defining 
	\begin{align}\label{eq:heisenberg}
		\hbr(t) &= \uu^\dagger  \hbr \uu,\\
		\frac{d}{dt} \hbr(t) &= \bm{\Omega} \bm{H} \hbr(t) \longrightarrow \hbr(t) = \underbrace{\exp(\bm{\Omega} \bm{H} t)}_{\bm{S}} \bm{r}.
	\end{align}
	It is straightforward to verify that the matrix $\bm{S}$ is symplectic, satisfying $\bm{S} \bm{\Omega} \bm{S}^T = \bm{\Omega}$.
	
	One may invoke Schur's lemma (cf. Appendices A and B of Ballantine~\cite{ballentine2014quantum}) to argue that any sequence of unitary operations $\hat{\mathcal{V}}$ that achieves
	$\hat{\mathcal{V}}^\dagger \bm{r} \hat{\mathcal{V}} = \bm{S} \bm{r}$ must be equal to $\uu$ up to a phase. Typically, this sequence is obtained by combining passive operations, which do not change the total photon number, and active operations in the form of squeezing\footnote{The precise way to obtain this sequence using the Bloch-Messiah decomposition of the symplectic matrix associated with the unitary transformation $\uu$ is discussed in Sec. 4.1 of  Ref.~\cite{kalajdzievski2021exact}. A self contained discussion of how to obtain the Bloch-Messiah decomposition of a symplectic matrix can be found in Ref.~\cite{houde2024matrix}.}. The method just described is successful at finding a suitable sequence of gates to decompose the action of a quadratic Hamiltonian but is of course unable to account for the phase in the argument just presented. This motivates the main problem addressed by this manuscript, how to obtain this phase which cannot be accounted for once one considers evolution in the Heisenberg picture Eq.~\eqref{eq:heisenberg}.
	
	To give an example of when this phase is non-trivial consider the single-mode quadratic-phase gate
	\begin{align}
		\hat{P}(t) = \exp(i t \hat{q}^2  / 2\hbar),
	\end{align}
	from which we read $\bm{H} = \left(\begin{smallmatrix}
		-1 & 0 \\ 0 & 0    
	\end{smallmatrix} \right)$ and find
	\begin{align}
		\bm{S} &= \exp(\bm{\Omega} \bm{H} t) = \begin{pmatrix}
			1 & 0 \\
			t & 1
		\end{pmatrix}\\& = \bm{R}(\theta) [e^r \oplus e^{-r}] \bm{R}( \tfrac{\pi}{2} - \theta), \ \bm{R}(\theta) =
		\begin{pmatrix}
			\cos \theta &-\sin \theta \\
			\sin \theta & \cos \theta
		\end{pmatrix}, \nonumber 
	\end{align}
	where the parameters above are related to $t\geq0$ as follows  $\sinh r = t/2, \  \theta = \tan^{-1} e^{r}$ (cf. Eq. 41 of Ref.~\cite{kalajdzievski2021exact}).
	The Bloch-Messiah decomposition above gives the following sequence in Fock space
	\begin{align}\label{eq:put}
		\hat{\mathcal{V}} &= \exp\left( i \theta \had \ha \right) \exp(\tfrac{r}{2}(\ha^{\dagger 2} - \ha^2)) \exp\left( i (\tfrac{\pi}{2} - \theta) \had \ha  \right).
	\end{align}

	It is easy to see that the equation above does not correctly capture the phase of the gate as
	\begin{align}
		\braket{0|\hat{\mathcal{V}}|0} = \frac{1}{\sqrt{\cosh r}} = \frac{1}{\sqrt[4]{1+\tfrac{t^2}{4}}} \in \mathbb{R}^+,
	\end{align}
	while for the actual gate one can use that $\braket{0|q} = \frac{1}{\sqrt[4]{\pi \hbar}} \exp(-q^2 / (2 \hbar))$ and that the (single-mode) identity operator can be resolved in terms of eigenkets of the position operator $\hat{\mathds{1}}\ = \int_{-\infty}^{\infty} dq \ket{q}\bra{q}$ to find 
	\begin{align}\label{eq:qphase}
		\braket{0|\hat{P}(t)|0} &= \int_{-\infty}^{\infty} dq \braket{0|P(t)|q} \braket{q|0}\nonumber \\
		&=\int_{-\infty}^{\infty} dq \braket{0|q} \exp\left( i\frac{t}{2 \hbar} q^2 \right) \braket{q|0} \nonumber \\
		&=\int_{-\infty}^{\infty} dq  \exp\left( i\frac{t}{2 \hbar} q^2 \right) |\braket{q|0}|^2  \nonumber \\
		&= \frac{1}{\sqrt{1 - i \tfrac{t}{2}}} \in \mathbb{C}. 
	\end{align}
	We see that the phase we are after is also the phase associated with the vacuum-to-vacuum amplitude of the unitary in Eq.~\eqref{eq:unit}, and we can write $\sqrt[4]{1+\tfrac{t^2}{4}}/\sqrt{1 - i \tfrac{t}{2}}$ as the correction to the decomposition in Eq.~\eqref{eq:put}.
	
	In the rest of the manuscript we show how the amplitude and phase
	\begin{align}
		\amp = \braket{\bm{0}| \exp\left(- \tfrac{i}{2\hbar} \hbr \bm{H}  \hbr t \right)| \bm{0}}, \  e^{i\varphi} = \amp/|\amp|,
	\end{align}
	can be obtained.
	Note that the absolute value of this  amplitude, the vacuum-to-vacuum \emph{probability}, is easy to find. To see this simply note 
	\begin{align}
		|\amp|^2 =  \bra{\bm{0}} \uu \ket{\bm{0}}  \bra{\bm{0}} \uu ^\dagger \ket{\bm{0}}.
	\end{align}
	This can be interpreted as the fidelity between two pure Gaussian states, namely $\uu \ket{\bm{0}}  \bra{\bm{0}} \uu ^\dagger$ and $ \ket{\bm{0}}  \bra{\bm{0}}$, for which we can use phase-space methods to obtain the probability~\cite{brask2021gaussian}
	\begin{align}\label{eq:vacprob}
		|\amp|^2 = \frac{1}{\sqrt{\det\left[ \frac{1}{2} (\bm{S} \bm{S}^T + \mathbb{I}_{2M}) \right]}}.
	\end{align}

	\section{Symplectically-diagonalizable Hamiltonians}\label{sec:Hpos}
	In this section we consider how to obtain the vacuum-to-vacuum amplitude of Hamiltonians that can be symplectically diagonalized. If the Hamiltonian matrix ${\bm{H}}$ can be diagonalized by a symplectic matrix $\Stwo$
	then there exist a Gaussian unitary operator $\TT$ such that $\TT^\dagger \hat{\bm{r}} \TT = \Stwo \hat{\bm{r}}$ and then
	\begin{align}\label{eq:factor}
		&	\bra{\bm{0}} \exp\Big(- \tfrac{i}{2 \hbar}  \hbr^T \underbrace{ \Stwo^T  [\bm{\mu}\oplus \bm{\nu}]  \Stwo }_{\equiv \bm{H}} \hbr  t \Big)\ket{\bm{0}} \\
		&=\bra{\bm{0}} \exp\left(- \tfrac{i}{2 \hbar} \TT^\dagger \hbr^T  [\bm{\mu}\oplus \bm{\nu}]   \hbr  \TT  t \right)\ket{\bm{0}}\nonumber\\
		& =  \bra{\bm{0}}\TT^\dagger \exp\left(- \tfrac{i}{2 \hbar} \hbr^T  [\bm{\mu}\oplus \bm{\nu}]  \hbr t \right) \TT \ket{\bm{0}}  = \bra{\bm{0}}\TT^\dagger \uu_1 \TT \ket{\bm{0}}, \nonumber
	\end{align}
	where
	\begin{align}
		\uu_1 = \exp\left(- \tfrac{i}{2 \hbar} \hbr^T  [\bm{\mu}\oplus \bm{\nu}]  \hbr t \right),
	\end{align}
	is a unitary generated by the diagonal Hamiltonian matrix  $\bm{H}_1= \bm{\mu}\oplus \bm{\nu}$.
	As it turns out if this decomposition is possible, we will show below that one can write a simple expression for the vacuum-to-vacuum amplitude in Eq.~\eqref{eq:factor}. To achieve this goal, in the following subsections we comment on conditions for the symplectic diagonalization of a real symmetric matrix, show how to obtain the phase associated with a single-mode Hamiltonian and then show how to combine this information together with the composition rules for Gaussian unitaries based on the Bargmann representation (see Appendices~\ref{app:bargmann} and~\ref{app:gaussian}) to obtain the net phase in terms of the symplectic $\Stwo$ and the diagonal elements of $ \bm{\mu}\oplus \bm{\nu}$. Note that the Bargmann formalism can also be used to calculate the expectation of a Gaussian unitary over any (mixed or pure) Gaussian state if the vacuum-to-vacuum amplitude is known. This is further discussed in Appendix~\ref{app:mixed_gaussians}.
	
	\subsection{Symplectic diagonalization}
	It would be desirable to always be able to find a symplectic diagonalization of a symmetric matrix. Conditions under which this is possible are considered in many references, for example Appendix 6 of the book by Arnol'd~\cite{arnol2013mathematical}, Chapter 3 of the book by Blaizot and Ripka~\cite{blaizot1986quantum} and Theorem 21.5.3 of the book by H\"{o}rmander~\cite{hormander2007analysis}.

	If the matrix ${\bm{H}}$ is PD, then one is guaranteed to obtain a symplectic decomposition by Williamson's theorem~\cite{williamson1937normal} and in this case, one finds that $\mu_i = \nu_i > 0$. Numerically, this decomposition can be obtained by performing a real Schur decomposition of an anti-symmetric matrix, which is numerically stable and can be performed using standard linear algebra packages. The general case requires the Jordan normal form of $\bm{\Omega} {\bm{H}}$~\cite{arnol2013mathematical}; however the calculation of this normal form is numerically unstable (see Sec. 7.6.5 of Golub and van Loan~\cite{golub2013matrix}).
	
	Despite these potential numerical difficulties other authors have studied necessary and/or sufficient conditions for symplectic diagonalization of non-PD matrices to exist~\cite{mishra2024generalization,kamat2024simultaneous,pereira2021symplectic}.
	Pereira, Banchi and Pirandola~\cite{pereira2021symplectic} studied generalizations of the Williamson theorem where one still demands $\bm{\mu} = \bm{\nu}$ but does not demand all of the symplectic eigenvalues to be positive. They find that, when this generalized decomposition exists, it can be obtained by calculating the symplectic eigenvalues in the usual way (namely by considering the eigenvalues of $i\bm{\Omega} {\bm{H}}$) and then considering determinants of sub-matrices constructed from $\bm{H}$. Numerical experiments suggest that it may be possible to generalize these results to the case where for some of the modes one has $\mu_i = - \nu_i$~\cite{pereira2025priv}.

	\subsection{Single-mode Hamiltonians}
	To obtain the vacuum-to-vacuum amplitude of a single mode Hamiltonian we start by assuming without loss of generality that it can be diagonalized by a rotation matrix $\bm{R}(\theta)$ (which for a single mode is also symplectic) and write
	\begin{align}
		\bm{H} = \bm{R}^T(\theta) [\mu \oplus \nu] \bm{R}(\theta),
	\end{align}
	with $|\mu|\geq |\nu|$. In the case where $\nu =0$ we obtain the case of the quadratic-phase gate, discussed earlier. 
	The diagonalization above does not directly give us a way to obtain the vacuum-to-vacuum amplitude. However if we could factorize the matrix in terms of a diagonal matrix where the absolute values of the diagonal entries are equal then we could use the composition rules for Gaussian unitaries, since the cases with equal or opposite signs (but equal absolute values) in the diagonal correspond to rotation and squeezing, for which the vacuum-to-vacuum amplitude is well-known~\cite{miatto2020fast}. This sought after factorization is given by
	\begin{align}\label{eq:fact}
		[\mu \oplus \nu]= \begin{pmatrix}
			z & 0 \\ 0 & 1/z
		\end{pmatrix}
		\sqrt{|\mu \nu|} \sign \mu
		\begin{pmatrix}
			1 & 0 \\ 0 & \sign \mu \nu
		\end{pmatrix} \begin{pmatrix}
			z & 0 \\ 0 & 1/z
		\end{pmatrix},
	\end{align}
	with $z = \sqrt[4]{\frac{|\mu|}{|\nu|}}$. With the above results we identify
	\begin{align} 
		\Stwo &=   \begin{pmatrix}
			z & 0 \\ 0 & 1/z
		\end{pmatrix} \bm{R}(\theta),  \
		\bm{H}_1 = \sqrt{|\mu \nu|} \sign \mu 
		\begin{pmatrix}
			1 & 0 \\ 0 & \sign \mu \nu
		\end{pmatrix}.
	\end{align}
	Depending on the sign of $\mu \nu$ the unitary $\uu_1$ corresponds to a rotation gate or a squeezing gate.
	In Fock space we are left to calculate $\TT^\dagger \uu_1 \TT$ where the phase of $\uu_1$ is known. 
	Using the Bargmann formalism, as detailed in Appendix~\ref{app:gaussian}, we can consider the two cases separately and arrive to a single unified formula that also covers the phase gate
	\begin{align}
		\label{eq:singlemode}
		\amp = \frac{1}{\sqrt{\cos t \sqrt{\det {\bm{H}}} + i  t \frac{\tr {\bm{H}}}{2} \sinc t  \sqrt{\det {\bm{H}}}}}.
	\end{align}
	This single-mode result could also have been obtained using compositions rules for squeezing operators, as described in Appendix A of Ref.~\cite{bressanini2022noise}, disentangling theorems for single mode operators, as described in Appendix 5 of
	Barnett and Radmore~\cite{barnett2002methods} or the triple Gaussian integral methods from Refs.~\cite{dias2024classical,paraoanu2000fidelity}.

	\subsection{Multi-mode Hamiltonians}
	Once we have the phase associated to a single-mode Hamiltonian we can use it to write down the phase associated to a direct sum of them
	\begin{align}
		\uu_1 &=  \exp\left(-\tfrac{i}{2\hbar} \sum_{j=1}^M (\mu_j \hat{q}_j^2 + \nu_j \hat{p}_j^2)\right)\nonumber \\
		&=\bigotimes_{j=1}^M \exp\left(-\tfrac{i}{2\hbar} (\mu_j \hat{q}_j^2 + \nu_j \hat{p}_j^2)\right),
	\end{align}
	as $\prod_{j=1}^M c_j$ with 
	\begin{align}
		c_j &= 	\frac{1}{\sqrt{\cos t \sqrt{\mu_j \nu_j} + i  t \frac{ {\mu_j+\nu_j}}{2} \sinc t  \sqrt{\mu_j \nu_j}}}.
	\end{align}
	We can also easily calculate the associated symplectic of the diagonal Hamiltonian $\bm{H}_1$ as
	\begin{align}
		\bm{S}_1 =  \exp(\bm{\Omega} \bm{H}_1 t) = \exp(\bm{\Omega} [\bm{\mu} \oplus \bm{\nu}] t).
	\end{align}
	Note that  in general, we can demand that $|\mu_i| = |\nu_i|$ if $\mu_i \nu_i \neq 0$ in the equations above. The reason why this is possible is implicit in Eq.~\eqref{eq:fact}, as we can always make them equal by conjugating with a symplectic transformation that can be absorbed into $\Stwo$.

	Having the amplitude and symplectic of the mode-diagonal Hamiltonian and by virtue of the fact that the phase of $\TT$ cancels out the phase $\TT^\dagger$ in Eq.~\eqref{eq:factor}, we can use the Bargmann formalism (cf. Appendix~\ref{app:gaussian}) once more to write the vacuum-to-vacuum amplitude of the Hamiltonian in Eq.~\eqref{eq:factor} as
	\begin{align}\label{eq:pnr}
		\amp =& \braket{\bm{0}|\exp\left(-\tfrac{i}{2 \hbar} \hat{\bm{r}}^T {\bm{H}} \hat{\bm{r}} t\right) |\bm{0}} \\
		=& \frac{\prod_{j=1}^M c_j}{\sqrt{\det\left( \mathbb{I}_{2M} -   \tfrac12\left[ \Stwo \Stwo^T -  \mathbb{I}_{2M}  \right] \left[\bm{K} -\mathbb{I}_{2M}\right]\right)}}
	\end{align}
	where 
	\begin{align}\label{eq:kdef}
		\bm{K} &= [\oplus_{j=1}^M c_j^2(1+\xi_j)] \oplus [\oplus_{j=1}^M c_j^2(1-\xi_j)],\\
		\xi_j &= -\tfrac{i}{2} (\mu_j -\nu_j) t \sinc t \sqrt{\mu_j \nu_j}.
	\end{align}
	If either the symplectic $\bm{S}_1$ associated with the diagonal Hamiltonian $\bm{H}_1$ or the diagonalizing symplectic $\Stwo$ is an orthogonal matrix we can further simplify the expression above.
	
	The first case is equivalent to $\bm{\mu} = \bm{\nu}$, as it happens when, for example, $\bm{H}$ is positive-definite, and then
	\begin{align}
		c_j = e^{-i \mu_j t /2}, \  \xi_j = 0, \ \bm{K} = [\oplus_{j=1}^M c_j^2] \oplus [\oplus_{j=1}^M c_j^2].
	\end{align}
	This case can be used to calculate the photon-number moment generating function of a Gaussian state with covariance matrix $\tfrac{\hbar}{2}\Stwo \Stwo^T$ (cf. Eq.~65 of Bulmer et al.~\cite{bulmer2024simulating} or Eq.~30 of Cardin et al.~\cite{cardin2024photon}).

	In the second case, the amplitude is the product of the single-mode amplitudes $\prod_{j=1}^M c_j$. Note that if a matrix is symplectic and orthogonal it can be written as
	\begin{align}
		\Stwo = \begin{pmatrix}
			\Re[\bm{U}] &  -\Im[\bm{U}] \\
			\Im[\bm{U}] & \Re[\bm{U}]
		\end{pmatrix},
	\end{align}
	with $\bm{U}$ a unitary matrix and $\Re[\bm{U}]$ and $\Im[\bm{U}]$ its real and imaginary parts repectively.
	In terms of this unitary we can consider the following special cases:
	
	--- For \emph{active} Hamiltonians one has
	\begin{align}
		\bm{\mu} = -\bm{\nu}, \ \frac{\hat{\mathcal{H}}}{2\hbar} = \sum_{kl} \left( f_{kl} \hat{a}_k^\dagger \hat{a}^\dagger_l + \text{H.c.}\right), \ \bm{f} = \bm{U}^\dagger \bm{\mu} \bm{U}^* ,
	\end{align}
	and then the vacuum-to-vacuum amplitude is simply $\sqrt{\prod_{j=1}^M \sech \mu_j t}$. Note that since $\bm{f} = \bm{f}^T$ we can always find $\bm{U}$ and $\bm{\mu}$ via a Takagi-Autonne decomposition of $\bm{f}$.
	
	--- For \emph{passive} Hamiltonians one  has
	\begin{align}
		\bm{\mu} = \bm{\nu}, \ \frac{\hat{\mathcal{H}}}{2\hbar} = \sum_{kl} \omega_{kl} \hat{a}_k^\dagger \hat{a}_l +\frac{\tr \bm{H}}{4}, \ \bm{\omega} = \bm{U}^\dagger \bm{\mu} \bm{U} ,
	\end{align}
	and then the vacuum-to-vacuum amplitude is simply $\exp(- i  t \tr \bm{H}/4)$. Note that since $\bm{\omega} = \bm{\omega}^\dagger$ we can always find $\bm{U}$ and $\bm{\mu}$ via an eigendecomposition of $\bm{\omega}$.
	
	--- Finally, \emph{quadrature-diagonal} Hamiltonians, which are a generalization of the phase-gate, are obtained when $\bm{\nu}=0$. In this case we can define a collection of generalized positions and momenta as
	\begin{align}
		\hat{\bm{x}} = \Re[\bm{U}] \hat{\bm{q}} -\Im[\bm{U}] \hat{\bm{p}}, \\
		\hat{\bm{y}} = \Im[\bm{U}] \hat{\bm{q}}  + \Re[\bm{U}] \hat{\bm{p}},
	\end{align}
	and write the Hamiltonian solely in terms of the generalized positions
	\begin{align}
		\bm{\nu} = 0, \frac{\hat{\mathcal{H}}}{2\hbar} = \sum_{k=1}^M \frac{\mu_k}{2\hbar} \hat{x}_k^2,
	\end{align}
	and obtain the vacuum-to-vacuum amplitude as $c = \prod_{k=1}^M\frac{1}{\sqrt{1 + \mu_j t /2}}$. Two important gates are in this category, the continuous-variable controlled not and controlled phase gates
	\begin{align}
		\hat{\text{C}}_X(t) = \exp\left(-i \tfrac{t}{\hbar} \hat{q}_1 \hat{p}_2 \right) ,
		\hat{\text{C}}_Z(t) = \exp\left(i \tfrac{t}{\hbar} \hat{q}_1 \hat{q}_2 \right),
	\end{align}
	which have $\mu_1 = -\mu_2 = 1$ and thus their vacuum-to-vacuum amplitude is $ 1/\sqrt{1+t^2/4} \in \mathbb{R}^+$.

	\section{Time-dependent Hamiltonians}\label{sec:general}
	In this section we consider time-dependent Hamiltonians. The Heisenberg equation of motion of the quadratures now generalizes to
	\begin{align}\label{eq:heistd}
		\frac{d}{dt} \hbr(t) &= \bm{\Omega} \bm{H}(t) \hbr(t)\\
		&\longrightarrow \hbr(t) = \underbrace{\vec{\mathfrak{T}}\exp\left(\int_{0}^t dt' \bm{\Omega}  \bm{H}(t')\right)}_{\bm{S}} \bm{r}. \nonumber
	\end{align}
	Here we have introduced the time-ordering operator $\vec{\mathfrak{T}}$ ~\cite{shankar2012principles}.
	We claim that the Heisenberg propagator is still a symplectic matrix by simply noticing that the time-order exponential can be interpreted as a trotterized evolution~\cite{trotter1959product}
	\begin{align}\label{eq:heistd1}
		&\vec{\mathfrak{T}}\exp\left(\int_{0}^t dt' \bm{\Omega} \bm{H}(t')\right) =\\ &\lim_{N\to\infty} e^{\bm{\Omega} \bm{H}(t_N) \Delta t} e^{\bm{\Omega} \bm{H}(t_{N-1}) \Delta t} \ldots e^{\bm{\Omega} \bm{H}(t_1) \Delta t} e^{\bm{\Omega} \bm{H}(t_0) \Delta t}, \nonumber
	\end{align}
	with $t_i = i \Delta t$ for $i \in [0,1,\ldots,N ]$ and $\Delta t = t/(N+1)$. Since each $e^{\bm{\Omega} \bm{H}(t_i) \Delta t}$ is a symplectic matrix and these form a group then the product above is also a symplectic matrix. For numerical purposes, one can approximate the infinite product above by a finite one, or use perturbation theory. If one wants to respect the symplectic nature of the problem it is necessary to use the Magnus expansion~\cite{blanes2009magnus} as the Dyson series (cf. e.g. Sec. 5.7. of Sakurai and Napolitano~\cite{sakurai2017modern}) will not respect this structural property of the dynamics~\cite{quesada2022beyond}. 
	
	Just like for the time-independent case, the phase of the transformation cannot be obtained from the Heisenberg picture. To find the phase for an arbitrary time-dependent Hamiltonian we would like to express $\hat{\mathcal{H}}$ in terms of ladder operators. After some algebra we find
	\begin{align}
		&\hat{\mathcal{H}}(t) =h(t) + :\hat{\mathcal{H}}(t):\\
		&:\hat{\mathcal{H}}(t): =%
		\tfrac{1}{2} \sum_{i,j=1}^M \underbrace{[E_{ij}(t)+G_{ij}(t) + i (F_{ji}(t)-F_{ij}(t))]}_{\equiv 2 \omega_{ij}(t)} a_i^\dagger a_j \nonumber \\
		&+ \tfrac{1}{4} \sum_{i,j=1}^M \underbrace{[E_{ij}(t)-G_{ij}(t) - i (F_{ij}(t)+F_{ji}(t))]}_{\equiv 4f_{ij}(t)} a_i^\dagger a_j^\dagger + \text{H.c.},\nonumber
	\end{align}
	where we wrote $h(t)=\tfrac14 \tr \bm{H}(t)$,  $\bm{H}(t) = \left(\begin{smallmatrix}
		\bm{E}(t) & \bm{F}(t) \\ \bm{F}^T(t) & \bm{G}(t)
	\end{smallmatrix} \right)$, with $\bm{E}(t) = \bm{E}(t)^T$, $\bm{G}(t) = \bm{G}(t)^T$ and $\bm{f}(t) = \bm{f}(t)^T$, $\bm{\omega}(t) = \bm{\omega}(t)^\dagger$ and introduced the normal-ordered operator $:\hat{\mathcal{H}}:$.
	We can then write
	\begin{align}\label{eq:ladder}
		\hat{\mathcal{H}}=  h(t) + \sum_{ij} \omega_{ij}(t)a_i^\dagger a_j  +\sum_{ij}  f_{ij}(t) a_i^\dagger a_j^\dagger  + \text{H.c.}
	\end{align}
	
	In Ref.~\cite{ma1990multimode} Ma and Rhodes consider time-dependent quadratic bosonic Hamiltonians like the one above (with extra linear terms of the form $ \sum_{i=1}^M  g_i(t) \had +\text{H.c.}$ that we will ignore).
	In their Eq. 4.8 they write a normal-ordered expression for the unitary evolution operator associated to this Hamiltonian as
	\begin{align}\label{eq:normalma}
		& \exp(A(t)) \exp\left[\sum_{i=1}^M B_i(t) \had_i + \sum_{ij=1}^M C_{ij}(t)\had_i \had_j \right] \\
		&\times :\exp\left(\sum_{ij=1}^M D_{ij}(t) \had_i \ha_j \right):\nonumber \\ &\times \exp\left[\sum_{i=1}^M E_i(t) \ha_i + \sum_{ij=1}^M F_{ij}(t)\ha_i \ha_j \right]. \nonumber
	\end{align}
	Since we do not consider linear terms in the Hamiltonian $g_i = 0 $ it follows that $E_i=B_i=0$.
	By taking the vacuum expectation value in the equation above and noticing that this expression is in normal order we can immediately conclude  $\amp = \exp(A(t))$; Ma and Rhodes show that this quantity only depends on the matrix $\bm{C}(t)$ and the scalar $h(t) = \tfrac14 \tr \bm{H}(t)$ as follows (Eq. 4.8a of Ref.~\cite{ma1990multimode})
	\begin{align}\label{eq:phase}
		\amp = \exp\left[-i \int_0^t dt' \left\{  \tfrac{1}{4} \tr \bm{H}(t)  + 2 \tr[\bm{f}^\dagger(t) \bm{C}(t')] \right\} \right],
	\end{align}	
	where $\bm{C}(t)$ satisfies the Riccati equation (Eq. 4.8c of Ref.~\cite{ma1990multimode})
	\begin{align}\label{eq:ricatti}
		i \frac{d}{dt} \bm{C}(t) = 4 \bm{C}(t) \bm{f}^\dagger(t) \bm{C}(t) + 2 \bm{\omega}(t) \bm{C}(t) + \bm{f}(t),
	\end{align}
	with initial value $\bm{C}(0)=\bm{0}_M$.
	We can write this equation in a more convenient way by introducing $\bm{\chi}(t) = 2\bm{C}(t)$ to find
	\begin{align}
		\frac{d}{dt} \bm{\chi}(t) = -i 2 \bm{\chi}(t) \bm{f}^\dagger(t) \bm{\chi}(t) -2 i  \bm{\omega}(t) \bm{\chi}(t) - 2i\bm{f}(t),
	\end{align} 
	This matrix-valued nonlinear differential equation can be solved using Radon's lemma~\cite{radon1928problem} (cf. Theorem 3.1.1. of Ref.~\cite{abou2012matrix}). First we form the linear system
	\begin{align}\label{eq:asslinear}
		\frac{d}{dt} \begin{pmatrix} \bm{R}(t) \\ \bm{S}(t)\end{pmatrix} &= i \underbrace{\begin{pmatrix} \om &  2 \bm{f}^\dagger(t) \\
				-2  \bm{f}(t) & -2  \bm{\omega}(t) \end{pmatrix}}_{\equiv \bm{\Lambda}(t)}  \begin{pmatrix} \bm{R}(t) \\ \bm{S}(t)\end{pmatrix} ,
	\end{align}
	with initial conditions	$\bm{R}(0) = \mathbb{I}_M$ and $\bm{S}(0)=\bm{0}_M$ (giving $\bm{C}(0) = \bm{\chi}(0)=\bm{0}_M$). The solution of  this linear system can now be written as $		\left(\begin{smallmatrix} \bm{R}(t) \\ \bm{S}(t)\end{smallmatrix} \right) = \vec{\mathfrak{T}} \exp\left[ i \int_0^t dt' \bm{\Lambda}(t')   \right] \left( \begin{smallmatrix} \mathbb{I}_M \\ \bm{0}_M \end{smallmatrix}\right),$
	and from it we obtain $\bm{\chi}(t) = \bm{S}(t) \bm{R}^{-1}(t)$ and the amplitude
	\begin{align}\label{eq:quad}
		\amp = \exp\left[   -i \int_0^t dt' \left\{ \tfrac{1}{4} \tr \bm{H}(t')  +\tr[\bm{f}^\dagger(t') \bm{S}(t') \bm{R}^{-1}(t')] \right\} \right].
	\end{align}
	Note that just like the case of the Heisenberg propagator, we can obtain the propagator  $\vec{\mathfrak{T}} \exp\left[ i \int_0^t dt' \bm{\Lambda}(t')   \right]$ by trotterization, although there are more sophisticated methods for the numerical integration of the Riccati equation~\cite{behr2019invariant}.
	
	The results just derived can also be employed for time-independent Hamiltonians, which can be helpful if a symplectic diagonalization of the Hamiltonian matrix is not known; in this case we can drop the time-ordered operator $\vec{\mathfrak{T}}$ and replace $\int_0^t dt' \to t $ in the exponential. 	
	One can evaluate the argument of the integral leading to the amplitude in Eq.~\eqref{eq:quad} with relative ease numerically: assuming that the matrix $\bm{\Lambda}$ can be diagonalized as $\bm{\Lambda} = \bm{U} \bm{\lambda} \bm{U}^{-1}$  with $\bm{\lambda}$ a diagonal matrix then we can immediately compute 
	$\exp(i \bm{\Lambda} t) = \bm{U} \exp( i \bm{\lambda} t) \bm{U}^{-1}$. If the diagonalization of the matrix $\bm{\Lambda}$ is known, calculating $\exp(i \bm{\Lambda} t) $ requires only matrix multiplication, and thus one can evaluate the integrand of Eq.~\eqref{eq:quad}. Once an efficient blackbox exists  for evaluation of the integrand then one can perform the integration with any numerical integration method. The success of these methods will depend on whether the integrand is highly oscillatory or well-behaved.

	\section{Final remarks}\label{sec:final}
	We have analyzed the vacuum-to-vacuum amplitude associated with unitaries generated by quadratic bosonic Hamiltonians. For Hamiltonians with positive-definite matrices, or that are symplectically diagonalizable, we derived a closed-form expression for this amplitude by leveraging the Bargmann representation of Gaussian unitaries.
	
	For more general quadratic Hamiltonians, we extended the seminal approach of Ma and Rhodes~\cite{ma1990multimode}, showing that the problem reduces to solving a system of coupled linear differential equations. These equations encode the nonlinear Riccati dynamics governing the amplitude's evolution and can be integrated numerically with polynomial cost in the number of modes. This provides a practical and scalable method for recovering the phase of arbitrary multimode Gaussian unitaries without resorting to infinite-dimensional representations.
	
	Finally, our approach immediately generalizes to time-dependent Hamiltonians. The differential equation~\eqref{eq:ricatti} is valid for Hamiltonians that are time dependent; it can be solved by trotterization~\cite{trotter1959product,suzuki1976generalized} (or more sophisticated methods~\cite{behr2019invariant}) and used to integrate Eq.~\eqref{eq:phase} and obtain the vacuum-to-vacuum amplitude of time-dependent Hamiltonians. Indeed, this is precisely how Heisenberg dynamics (Eq.~\eqref{eq:heistd} and Eq.~\eqref{eq:heistd1},) is solved-for with time-dependent Hamiltonians, an strategy that is heavily employed in the context of quantum nonlinear optics~\cite{quesada2020theory,helt2020degenerate,kopylov2025theory,houde2024perfect,kim2024simulation,kim2025simulating}.
	
	Our findings are of fundamental interest, as they allow for the calculation of (perhaps) the last element needed to fully characterize a Gaussian unitary in Fock space.
	They could also be of interest in more applied settings where it is possible to engineer the coupling of bosonic modes (electromagnetic~\cite{bourassa2021fast} or mechanical~\cite{wanjura2023quadrature}, for example) to qubits.
	The phase discussed in this work will dictate the type of hybrid entangled states that can be generated between discrete and continuous degrees of freedom~\cite{andersen2015hybrid}.

	\section*{Acknowledgements}
	The author thanks J.E. Sipe, C. Vendromin, D. Patel, H. de Guise. R. Garc\'ia-Patr\'on, R. Alexander, H. K. Mishra, L. Banchi, J. Pereira and F. Miatto for insightful discussions, J. Mart\'inez-Cifuentes, M. Houde, O. Solodovnikova  and P. Blinova for providing feedback on the manuscript and the MEIE du Qu\'ebec and NSERC of Canada for financial support.
	\appendix
	
	\section{Bargmann representation and the composition of bosonic quadratic unitaries}\label{app:bargmann}
	In this and the rest of the appendices we partition matrices and vectors as follows,
	\begin{align}
		\bm{A} = \begin{pmatrix}
			\bm{B} & \bm{C} \\ \bm{C}^T & \bm{D}
		\end{pmatrix},  \ \bm{b} = \begin{pmatrix} \bm{c} \\ \bm{d} \end{pmatrix},
	\end{align}
	when writing the Bargmann representation of a Gaussian unitary. In the equations above $\bm{A} = \bm{A}^T$ is of even size and $\bm{B}$, $\bm{C}$ and $\bm{D}$ are of the same dimension (half the dimension of $\bm{A}$). Similarly, $\bm{c}$ and $\bm{d}$ are of the same dimensions, half of $\bm{b}$. The names used for these quantities follow the conventions of Ref.~\cite{yao2024riemannian}. Unfortunately, in Ma and Rhodes~\cite{ma1990multimode} variables with the same name are used  and we continue to use the same names in the main text (cf. Eq.~\eqref{eq:normalma}). We alert the reader that there might be a small number of variable name collisions between the variables in Sec.~\ref{sec:general} and the appendices, however their meaning should (hopefully) be easy to tell from the context in which they appear.
	
	We consider a general unitary $\uu$ generated by a quadratic Hamiltonian (that may contain linear terms) and write its Bargmann representation which is the matrix element between two coherent states as~\cite{yao2024riemannian,motamedi2025stellar}
	\begin{align}
		\braket{\bm{\alpha}^*|\uu|\bm{\beta}} =& c \exp\left(-\tfrac12 [||\bm{\alpha}||^2 + ||\bm{\beta}||^2 ] \right) \\
		& \times  \exp\left( \bm{b}^T [\bm{\alpha} \oplus \bm{\beta}] \right) \nonumber \\
		& \times \exp\left( \tfrac12 [\bm{\alpha} \oplus \bm{\beta}]^T \bm{A} [\bm{\alpha} \oplus \bm{\beta}] \right). \nonumber
	\end{align}	%
	We recall that a coherent state $\ket{\bm{\beta}}$ with $\bm{\beta} \in \mathbb{C}^M$ is obtained by applying the Weyl (or displacement) operator
	\begin{align}
		\W(\bm{r}) =\exp\left( \tfrac{i}{\hbar} \hat{\bm{r}}^T \bm{\Omega} \bm{r}\right) = \exp\left( - \tfrac{i}{\hbar} {\bm{r}}^T \bm{\Omega} \hat{\bm{r}}\right),
	\end{align}
	with parameter $\bm{r} = \sqrt{2\hbar} \left(  \Re[\bm{\beta}] \oplus \Im[\bm{\beta}] \right)$ to the vacuum state $\ket{\bm{0}}$. %
	
	If we write $\uu^\dagger \hat{\bm{r}} \uu = \bm{S} \hat{\bm{r}} +\bar{\bm{r}}$ (we split $\bar{\bm{r}} =\bar{\bm{q}} \oplus \bar{\bm{p}}$) and write the Bloch-Messiah decomposition of the symplectic matrix as
	\begin{align}
		\bm{S} &=
		\begin{bmatrix}
			\Re(\bm{U}) & -\Im(\bm{U})\\
			\Im(\bm{U}) & \Re(\bm{U})
		\end{bmatrix} 
		\begin{bmatrix}
			e^{-\bm{r}} & \bm{0}_M \\
			\bm{0}_M & e^{\bm{r}}
		\end{bmatrix}
		\begin{bmatrix}
			\Re(\tilde{\bm{U}}) & -\Im(\tilde{\bm{U}})\\
			\Im(\tilde{\bm{U}}) & \Re(\tilde{\bm{U}})
		\end{bmatrix} ,
	\end{align}
	then we can write
	\begin{align}\label{eq:Auni}
		\bm{A} =& \begin{bmatrix}
			\bm{U} & \bm{0}_M \\
			\bm{0}_M & \tilde{\bm{U}}^T
		\end{bmatrix}
		\begin{bmatrix}
			-\tanh \bm{r} & \sech \bm{r} \\
			\sech \bm{r} & \tanh \bm{r} \\
		\end{bmatrix}
		\begin{bmatrix}
			\bm{U} & \bm{0}_M \\
			\bm{0}_M & \tilde{\bm{U}}^T
		\end{bmatrix}^T = \bm{A}^T \nonumber \\
		\bm{b} =&  \begin{pmatrix} \bar{\bm{q}} + i \bar{\bm{p}} + \bm{U} \tanh \bm{r} \ \bm{U} ^T \left( \bar{\bm{q}} - i \bar{\bm{p}} \right) \\
			- \tilde{\bm{U}}^T \sech \bm{r} \bm{U}^\dagger \left( \bar{\bm{q}} + i \bar{\bm{p}} \right) \end{pmatrix},
	\end{align}
	and $c$ is the vacuum-to-vacuum amplitude that is the subject of the main text of this manuscript, and is the only quantity that is not uniquely determined by $\bm{S}$ and $\bar{\bm{r}}$. Note however that 
	\begin{align}\label{eq:bargc2}
		|c|^2 =& \sqrt{\det(\mathbb{I}_M - \bm{B} \bm{B}^*)} \times \\
		&\exp\left(-\tfrac{1}{4\hbar}\left[ ||\bar{\bm{r}}||^2 +  (\bar{\bm{q}}^T - i \bar{\bm{p}}^T) \bm{B} (\bar{\bm{q}} -i \bar{\bm{p}}) + \text{c.c.} \right] \right), \nonumber
	\end{align}
	which reduces to Eq.~\eqref{eq:vacprob} in the limit of zero displacements.

	When $\uu$ is the Fock space identity operator $\hat{\mathds{1}}$ one has 
	\begin{align}
		\bm{A} = \bm{X} = \begin{pmatrix} \bm{0}_M & \mathbb{I}_M \\  \mathbb{I}_M & \bm{0}_M \end{pmatrix},\  \bm{b}=0, \ c=1 . 
	\end{align} 
	If one composes two Gaussian unitaries to obtain a third one $\uu_3 = \uu_1 \uu_2$ and assumes the unitary $\uu_i$ has triple $\bm{A}_i , \bm{b}_i, c_i$ then the following composition rule holds
	\begin{align}
		\bm{A}_3 =& \bm{B}_1 \oplus \bm{D}_2 + \left\{ \bm{C}_1 \oplus \bm{C}_2^T \right\} \bm{\mathcal{Z}} \left\{\bm{C}_1^T \oplus \bm{C}_2 \right\},  \\
		\bm{b}_3 =& \bm{c}_1 \oplus  \bm{d}_2 +  [\bm{C}_1 \oplus \bm{C}_2^T] \bm{\mathcal{Z}} [\bm{d}_1 \oplus \bm{c}_2],\\
		c_3 =& \frac{c_1 c_2}{\sqrt{\det(\bm{\mathcal{Y}})}}   \exp\left( \tfrac12  [\bm{d}_1^T \oplus  \bm{c}_2^T ] \bm{\mathcal{Z}} [
		\bm{d}_1  \oplus \bm{c}_2]
		\right),
	\end{align}	
	with
	\begin{align}
		\bm{\mathcal{Y}} &= \mathbb{I}_M - \bm{B}_2 \bm{D}_1 ,\\
		\bm{\mathcal{Z}} &= \bm{\mathcal{Z}}^T = \begin{bmatrix}  -\bm{D}_1 & \mathbb{I}_M\\ \mathbb{I}_M & -\bm{B}_2
		\end{bmatrix}^{-1} = \begin{bmatrix} \bm{\mathcal{Y}}^{-1} \bm{B}_2 & \bm{\mathcal{Y}}^{-1} \\ [\bm{\mathcal{Y}}^T]^{-1} & \bm{D}_1 \bm{\mathcal{Y}}^{-1}\end{bmatrix}.
	\end{align}
	
	\section{Vacuum-to-vacuum amplitude of a unitary-conjugated unitary}\label{app:gaussian}
	We now use the results from Appendix~\ref{app:bargmann} to compute the vacuum-to-vacuum amplitude of 
	\begin{align}\label{eq:produu}
		\uu = \uu_2^\dagger \uu_1 \uu_2.
	\end{align}
	We assume without loss of generality that $\bm{b}_1 = \bm{b}_2 = 0$. A direct application of the the composition rules discussed in the previous section gives
	\begin{align}\label{eq:vactovacgen}
		c=&\frac{c_1 |c_2|^2}{\sqrt{\det \bm{\mathcal{Y}} \det (\mathbb{I}_M  - \bm{\mathcal{X}} \bm{B}_2^*)  }} 
	\end{align}
	with $\bm{\mathcal{Y}} = \mathbb{I}_M-\bm{B}_2 \bm{D}_1$ and $\bm{\mathcal{X}} =\bm{B}_1+\bm{C}_1 \bm{\mathcal{Y}}^{-1} \bm{B}_2 \bm{C}_1^T$ %
	To obtain the equation above we used the fact that if $\uu_2$ has the triple $\bm{A}_2,\bm{b}_2,c_2$ then $\uu_2^\dagger$ has the triple $\bm{X} \bm{A}_2^* \bm{X}, \bm{X}\bm{b}_2^*, c_2^*$. Moreover, note that to complete the calculation we only require the top left corner of $\bm{A}_2$, namely $\bm{B}_2$, since we can use Eq.~\eqref{eq:bargc2} to obtain $|c_2|^2$.
	We now show how the equation above can be written in a more convenient way. First note that
	\begin{align}\label{eq:partA}
		\det \bm{\mathcal{Y}} \det (\mathbb{I}_M  - \bm{\mathcal{X}} \bm{B}_2^*)  = \det\left( \begin{bmatrix}
			\bm{\mathcal{Y}} & \bm{B}_2 \bm{C}_1^T \\ \bm{B}_2^* \bm{C}_1 & \mathbb{I}_M -\bm{B}_2^* \bm{B}_1	
		\end{bmatrix} \right).
	\end{align}
	The correctness of the expression above can be seen by using Schur complements and Sylvester's determinant identity $\det(\mathbb{I} - \bm{A} \bm{B}) = \det(\mathbb{I} - \bm{B} \bm{A})$. We can now write the expression for $\bm{\mathcal{Y}}$ and insert a resolution of the identity $\bm{X} \bm{X} = \mathbb{I}_{2M}$ to write
	\begin{align}
		&\begin{bmatrix}
			\bm{\mathcal{Y}} & \bm{B}_2 \bm{C}_1^T \\ \bm{B}_2^* \bm{C}_1 & \mathbb{I}_M -\bm{B}_2^* \bm{B}_1	
		\end{bmatrix} \bm{X} \bm{X} \\
		&= \begin{bmatrix}
			\bm{B}_2 \bm{C}_1^T & \mathbb{I}_M - \bm{B}_2 \bm{D}_1  \\  \mathbb{I}_M -\bm{B}_2^* \bm{B}_1 & \bm{B}_2^* \bm{C}_1	
		\end{bmatrix}   \bm{X}. \nonumber
	\end{align}
	We will now add and subtract $[\bm{B}_2 \oplus \bm{B}_2^*] \bm{X}$ to obtain
	\begin{align}
		=&\begin{bmatrix}
			\bm{B}_2 \bm{C}_1^T & \mathbb{I}_M - \bm{B}_2 \bm{D}_1  \\  \mathbb{I}_M -\bm{B}_2^* \bm{B}_1 & \bm{B}_2^* \bm{C}_1	
		\end{bmatrix}  \bm{X} \\
		&+ [\bm{B}_2 \oplus \bm{B}_2^*] \bm{X} -  [\bm{B}_2 \oplus \bm{B}_2^*] \bm{X} \nonumber\\
		=&\left( \begin{bmatrix}
			\bm{B}_2 & \mathbb{I}_M \\
			\mathbb{I}_M & \bm{B}_2^*
		\end{bmatrix} - \begin{bmatrix} \bm{B}_2 & \bm{0}_M \\ \bm{0}_M & \bm{B}_2^* \end{bmatrix} \begin{bmatrix} \mathbb{I}_M - \bm{C}_1^T & \bm{D}^1 \\ \bm{B}_1 & \mathbb{I}_M - \bm{C}_1^T \end{bmatrix} \right) \bm{X} \nonumber 
	\end{align}
	To summarize, the determinant of the expression above is equal to Eq.~\eqref{eq:partA}. We now recall that
	\begin{align}
		|c_2|^2 = \sqrt{\det(\mathbb{I}_M -\bm{B}_2 \bm{B}_2^*)} = \frac{1}{\sqrt{\det\left( \bm{X} \begin{bmatrix}
					\bm{B}_2 & \mathbb{I}_M \\
					\mathbb{I}_M & \bm{B}_2^*
				\end{bmatrix}^{-1}\right)}}
	\end{align}
	which allows us to write
	\begin{align}
		&\frac{|c_2|^2}{\sqrt{\det \bm{\mathcal{Y}} \det (\mathbb{I}_M  - \bm{\mathcal{X}} \bm{B}_2^*)}} \\
		&=\frac{1}{\sqrt{\mathbb{I}_{2M} - \begin{bmatrix}
					\bm{B}_2 & \mathbb{I}_M \\
					\mathbb{I}_M & \bm{B}_2^*
				\end{bmatrix}^{-1} \begin{bmatrix}
					\bm{B}_2 & \bm{0}_M \\
					\bm{0}_M & \bm{B}_2^*
				\end{bmatrix} \begin{bmatrix} \mathbb{I}_M - \bm{C}_1^T & \bm{D}_1 \\ \bm{B}_1 & \mathbb{I}_M -  \bm{C}_1
				\end{bmatrix}
		}} .\nonumber 
	\end{align}
	At this point we recall that 
	\begin{align}
		\begin{bmatrix}
			\bm{B}_2 & \mathbb{I}_M \\
			\mathbb{I}_M & \bm{B}_2^*
		\end{bmatrix}^{-1} \begin{bmatrix}
			\bm{B}_2 & \bm{0}_M \\
			\bm{0}_M & \bm{B}_2^*
		\end{bmatrix} = \tfrac12 \bm{R} \bm{X} \left(\mathbb{I}_{2M} - \bm{S}_2\bm{S}_2^T \right) \bm{X}^T \bm{R}^\dagger.
	\end{align}
	where $\bm{S}_2$ is the symplectic associated with $\uu_2$, $\uu_2^\dagger \hat{\bm{r}} \uu_2 = \bm{S}_2 \hat{\bm{r}}$.
	We can then write the vacuum-to-vacuum amplitude of $\uu_2^\dagger \uu_1 \uu_2$ as
	\begin{align}\label{eq:simpler}
		&\frac{c_1}{\sqrt{\mathbb{I}_{2M} - \tfrac12  \bm{R} \bm{X} \left(\mathbb{I}_{2M} - \bm{S}_2\bm{S}_2^T \right) \bm{X}^T \bm{R}^\dagger  \begin{bmatrix} \mathbb{I}_M - \bm{C}_1^T & \bm{D}_1 \\ \bm{B}_1 & \mathbb{I}_M -  \bm{C}_1
				\end{bmatrix}
		}} \nonumber \\
		&=\frac{c_1}{\sqrt{\mathbb{I}_{2M} - \tfrac12   \left( \bm{S}_2\bm{S}_2^T  - \mathbb{I}_{2M}\right) \left(\bm{K} - \mathbb{I}_{2M} \right)
		}}
	\end{align}
	where we introduced
	\begin{align}
		\bm{K} &=  \bm{X}^T \bm{R}^\dagger  \begin{bmatrix}  \bm{C}_1^T & -\bm{D}_1 \\ -\bm{B}_1 &   \bm{C}_1
		\end{bmatrix} \bm{R} \bm{X} \\
		&= \tfrac12 \begin{bmatrix} \bm{B}_1 + \bm{D}_1 + \bm{C}_1 + \bm{C}_1^T  & i (\bm{D}_1-\bm{B}_1+\bm{C}_1 - \bm{C}_1^T) \\
			i (\bm{D}_1-\bm{B}_1-\bm{C}_1 + \bm{C}_1^T) & \bm{C}_1 + \bm{C}_1^T -\bm{B}_1 - \bm{D}_1
		\end{bmatrix}
	\end{align}
	This formula is very convenient to apply for cases where $\uu_1 = \exp\left(\tfrac{i}{2\hbar} \sum_{j=1}^M (\mu_i \hat{q}_i^2 + \nu_i \hat{p}_i^2)\right)$ as in this case
	\begin{align}
		\bm{B}_1 &= \bm{D}_1 = \oplus_{i=1}^M c_i^2 \xi_i \\
		\bm{C}_1 &= \bm{C}_1^T = \oplus_{i=1}^M c_i^2 \\
		\bm{K} &= [\bm{C}_1 +\bm{B}_1] \oplus [\bm{C}_1 -\bm{B}_1]
	\end{align}
	where $c_j$ and $\xi_j$ are given in the main text.
	
	We can also use the result just derived for a single mode gate. We use
	\begin{align}
		\bm{S}_2 = [z \oplus 1/z], \ z=\sqrt[4]{\frac{\mu}{\nu}},
	\end{align}
	where now we need two cases depending on the relatives signs of $\mu$ and $\nu$.
	If (i) their signs are equal or (ii) they are opposite we have the following Hamiltonian matrix and $c$ and $\xi$ parameters
	\begin{align}\label{eq:simplerhamils}
		(i)\ &  \sqrt{\mu \nu} \sign{\mu} \mathbb{I}_2, c = e^{i t \sqrt{\mu \nu} \sign \mu  /2}, \xi = 0,\\
		(ii)\ & \sqrt{\mu \nu} \sign{\mu} \bm{\sigma}_Z, c = \sqrt{\sech t \sqrt{|\mu \nu|}}, \xi = -i \sinh t \sqrt{|\mu \nu|}.
	\end{align}
	When plugging these two sets of values into Eq~\eqref{eq:simpler} we obtain Eq.~\eqref{eq:singlemode} in the main text. Note that the values $c$ and $\xi$ are obtained \emph{solely} from the knowledge of the Hamiltonians $\sqrt{\mu \nu} \sign{\mu} \mathbb{I}_2$ and $ \sqrt{\mu \nu} \sign{\mu} \bm{\sigma}_Z$ ($\bm{\sigma}_Z$ is the usual Pauli-Z matrix) which correspond to a pure rotation and single-mode squeezing.

	\section{Extensions to mixed Gaussian states}\label{app:mixed_gaussians}
	In this section we consider how to obtain the phase of the quantity
	\begin{align}
		\tr \left[ \uu \hat{\varrho}(\bm{V},\bar{\bm{d}})\right] ,
	\end{align}
	assuming one knows the phase of $\braket{\bm{0}|\uu|\bm{0}}$. In the equation above it is assumed that $\uu$ is a Gaussian unitary and that $ \hat{\varrho}(\bm{V},\bar{\bm{r}})$ is a general (possibly-mixed) Gaussian state with covariance matrix $\bm{V}$ and vector of first-moments $\bar{\bm{r}}$. The Williamson decomposition~\cite{williamson1937normal} of a quantum covariance matrix $\bm{V}$ states that it can be split as
	\begin{align}\label{eq:split}
		\bm{V} = \vp+\vn,
	\end{align}
	where $\vp = \tfrac{\hbar}{2}\Stwo \Stwo^T$ is the covariance matrix of a pure state, $\Stwo$ is a symplectic matrix, and $\vn$ is positive semidefinite. In Hilbert space, this implies that a mixed Gaussian state with covariance matrix $\bm{V} = \vp+\vn$ and a vector of means $\bar{\bm{r}} = \bar{\bm{q}} \oplus \bar{\bm{p}} $ can be expressed as~\cite{wolf2004gaussian,serafini2023quantum}
	\begin{align}\label{eq:decomp} 
		\hat{\varrho} = \int d \bm{R} \  	p( \bm{R}) \ 
		\pr{\psi_{\bm{R}, \vp}},
	\end{align}
	where  
	\begin{align}\label{eq:mnpdf}
		p( \bm{R})&=  \frac{\exp\left( -\tfrac{1}{2} (\bm{R} - \bar{\bm{d}})^T  \ \vn^{-1} \ (\bm{R} - \bar{\bm{d}}) \right)}{\sqrt{\det(2 \pi  \vn)}},
	\end{align} 
	is the probability density function of a multivariate normal distribution with mean $\bar{\bm{d}}$ and covariance matrix $\vn$, and $\ket{\psi_{\bm{R}, \vp}}$ is a pure Gaussian state with vector of means $\bm{R}$ and covariance matrix $\vp$. We can write this pure state as $\ket{\psi_{\bm{R}, \vp}} = \hat{\mathcal{W}}(\bm{R}) \TT \ket{\bm{0}}$ up to a global phase; notice that this phase does not play any role in Eq.~\eqref{eq:decomp}. We can now write 
	\begin{align}
		\tr \left[ \uu \hat{\varrho}(\bm{V},\bar{\bm{d}})\right]  =&  \tr\left[ \uu \int d \bm{R} \  	p( \bm{R}) \ \pr{\psi_{\bm{R}, \vp}}\right] \nonumber \\
		=& \int d \bm{R} \  	p( \bm{R}) \tr\left[ \uu  
		\pr{\psi_{\bm{R}, \vp}}\right] \nonumber \\
		=& \int d \bm{R} \  	p( \bm{R}) \bra{\psi_{\bm{R}, \vp}} \uu  
		\ket{\psi_{\bm{R}, \vp}} \\
		=& \int d \bm{R} \  	 p( \bm{R}) \bra{\bm{0}}  \TT^\dagger \hat{\mathcal{W}}(\bm{R})^\dagger  \uu  
		\hat{\mathcal{W}}(\bm{R}) \TT \ket{\bm{0}} \nonumber
	\end{align}
	We now note that the quantum expectation value above can be calculated using the Bargmann formalism in Appendix~\ref{app:bargmann}. Moreover this expectation value will be a Gaussian function in the variable $\bm{R}$, whose average over the Gaussian distribution $p(\bm{R})$ can be easily obtained thus completing the calculation.

	\section{Linear terms}\label{app:linear}
	Here we study a generalization of the problem considered in the main text,
	\begin{align}
		\hat{\mathcal{Q}} = \exp\left[ -\tfrac{i}{\hbar} \left( \tfrac{1}{2} \hat{\bm{r}}^T \bm{H} \hat{\bm{r}} + \hat{\bm{r}}^T \bar{\bm{r}}\right)\right],
	\end{align}
	where $\bar{\bm{r}} \in \mathbb{R}^{2M}$. 
	It will be useful to %
	recall the  Heisenberg action of the displacement (Weyl) operator
	\begin{align}
		\W^\dagger(\bm{r}) \bm{\hat r} \W (\bm{r}) = \bm{\hat r}+\bm{r}.
	\end{align}
	
	To obtain a decomposition we first eliminate the linear part in the exponential by inserting resolutions of the identity in terms of $\W$ as follows
	\begin{align}
		\hat{\mathcal{Q}} =& \W(\br) \W^\dagger(\br) \hat{\mathcal{Q}} \W(\br) \W^\dagger(\br) \nonumber \\ 
		=&  \W(\br)\\
		& \times \exp\left[ -\tfrac{i}{\hbar} \left( \tfrac{1}{2} (\hat{\bm{r}} + \br)^T \bm{H} (\hat{\bm{r}} + \br) + (\hat{\bm{r}}+\br)^T \bar{\bm{r}}\right)\right] \nonumber \\
		&\times \W^\dagger (\br) \nonumber \\
		=&e^{-\frac{i}{2 \hbar} \br^T \bm{H} \br -\frac{i}{\hbar } \br^T \bar{\br}} \\
		&\W( \br ) \exp\left[ -\tfrac{i}{\hbar} \left( \tfrac{1}{2} \hat{\bm{r}}^T \bm{H} \hat{\bm{r}} +\hbr^T\{\bar{\br} +  \bm{H} \br \} \right) \right]  \W^\dagger(\br). \nonumber
	\end{align}
	If we pick as displacement 
	\begin{align}\label{choice}
		\br = -\bm{H}^{-1} \bar{\br}
	\end{align} then the term inside the curly braces in the last equation vanishes and we are left with an operator with a purely quadratic generator,
	\begin{align}
		\uu = \exp\left[ - \tfrac{i}{2 \hbar } \hat{\bm{r}}^T \bm{H} \hat{\bm{r}}  \right].
	\end{align}
	Moreover, under the choice of Eq.~\eqref{choice} one obtains
	\begin{align}
		\br^T \bm{H} \br +2 \br^T \bar{\br} = - \bar{\br}^T \bm{H}^{-1} \bar{\br}.
	\end{align}
	Putting everything together we have
	\begin{align}
		\hat{\mathcal{Q}} &= e^{\frac{i}{2 \hbar }  \bar{\br}^T \bm{H}^{-1} \bar{\br}} \W( \br ) \uu \W^\dagger( \br )  \\
		&=e^{\frac{i}{2 \hbar }  \bar{\br}^T \bm{H}^{-1} \bar{\br}} \W( \br ) \uu \W( \br )^\dagger \uu^\dagger  \uu.
	\end{align}
	We now recall Eq.~\eqref{eq:heisenberg} 
	\begin{align}
		\uu \W( \br )^\dagger \uu^\dagger &= \exp\left(  \tfrac{i}{\hbar} {\br}^T \bm{\Omega} \uu \hat{\bm{r}} \uu^\dagger\right) \\
		&=\exp\left(  \tfrac{i}{\hbar} {\br}^T \bm{\Omega} \bm{S}^{-1} \hat{\bm{r}} \right) \nonumber\\
		&=\exp\left(  \tfrac{i}{\hbar} {\br}^T \bm{S}^{T}\bm{\Omega}  \hat{\bm{r}} \right)\nonumber \\
		&=\W(-\bm{S} \bm{r}) = \W(\bm{S} \bm{H}^{-1} \bar{\bm{r}}), \nonumber 
	\end{align}
	where we used $\bm{S}^T \bm{\Omega} = \bm{\Omega}  \bm{S}^{-1}$.
	Now we need to simplify a product of two displacement operators for which we use
	\begin{align}
		\W(\bm{r}_1 + \bm{r}_2) e^{-\frac{i}{2 \hbar} \bm{r}_1^T \bm{\Omega} \bm{r}_2} = \W(\bm{r}_1) \W(\bm{r}_2),
	\end{align}
	to obtain
	\begin{align}
		&    \W( - \bm{H}^{-1} \bar{\br} )  \W(\bm{S} \bm{H}^{-1} \bar{\bm{r}}) \\
		&    = \W([\bm{S} - \mathbb{I}_{2M}] \bm{H}^{-1} \bar{\bm{r}}) 
		\exp( \tfrac{i}{2\hbar} \bar{\bm{r}}^T \bm{H}^{-1} \bm{\Omega} \bm{S} \bm{H}^{-1} \bar{\bm{r}} ).\nonumber 
	\end{align}
	The whole decomposition is now
	\begin{align}
		\hat{\mathcal{Q}} =& \exp\left( \tfrac{i}{2\hbar} \left[  \bar{\bm{r}}^T \bm{H}^{-1} \bm{\Omega} \bm{S} \bm{H}^{-1} \bar{\bm{r}} + \bar{\br}^T \bm{H}^{-1} \bar{\br} \right]  \right) \nonumber \\
		&\times \W([\bm{S} - \mathbb{I}_{2M}] \bm{H}^{-1} \bar{\bm{r}})  \times \uu.
	\end{align}
	
	Below we show that, even if $\bm{H}$ is non-invertible, this expression is still well behaved. Consider first the argument of the Weyl operator
	\begin{align}
		[\bm{S} - \mathbb{I}_{2M}] \bm{H}^{-1} =& \left[\sum_{n=0}^\infty \frac{(\bm{\Omega} \bm{H})^n}{(n+1)!} \bm{\Omega} \bm{H} \right] \bm{H}^{-1} \\
		=& \sum_{n=0}^\infty \frac{(\bm{\Omega} \bm{H})^n}{(n+1)!} \bm{\Omega} \equiv F_1(\bm{\Omega}\bm{H}) \bm{\Omega}, \nonumber
	\end{align}
	where $F_1(x) = \frac{e^x - 1}{x}$ is a well-behaved function of its argument. Now consider the phase
	\begin{align}
		&\bar{\br}^T \bm{H}^{-1} \bar{\br} +\bar{\bm{r}}^T \bm{H}^{-1} \bm{\Omega} \bm{S} \bm{H}^{-1} \bar{\bm{r}}  =  \bar{\br}^T \bm{H}^{-1} \bar{\br}  \\
		&\quad +\bar{\bm{r}}^T \bm{H}^{-1} \bm{\Omega} \left[ \mathbb{I}_{2M} + \bm{\Omega}\bm{H} + \sum_{n=0}^\infty \frac{(\bm{\Omega \bm{H}})^{n+2}}{(n+2)!} \right] \bm{H}^{-1} \bar{\bm{r}} \nonumber  
	\end{align}
	Note that the first term inside the square brackets will be multiplied by the  antisymmetric symplectic matrix and then will contract to zero when multiplied from the left and the right by the same vector (as this forms a symmetric tensor). The second term inside the square brackets cancels out exactly the first term of the equation above leaving us with the third of the square bracket that can be simplified to
	\begin{align}
	&	\bar{\bm{r}}^T \bm{H}^{-1} \bm{\Omega} \sum_{n=0}^\infty \frac{(\bm{\Omega \bm{H}})^{n+2}}{(n+2)!}  \bm{H}^{-1} \bar{\bm{r}}
	=-	\bar{\bm{r}}^T   \left[   \sum_{n=0}^\infty \frac{(\bm{\Omega \bm{H}})^{n}}{(n+2)!}   \right] \bm{\Omega} \bar{\bm{r}} \nonumber \\
	&=- \bar{\bm{r}}^T F_2(\bm{\Omega}\bm{H})  \bm{\Omega}  \bar{\bm{r}},	
	\end{align}
	where now $F_2(x) = \frac{e^x - x - 1}{x^2}$.
	Our final result is
	\begin{align}
		\hat{\mathcal{Q}} = e^{-\tfrac{i}{2 \hbar}  \bar{\bm{r}}^T  F_2[\bm{\Omega} \bm{H}] \bm{\Omega } \bar{\bm{r}} } \W(F_1[\bm{\Omega} \bm{H}] \bm{\Omega} \bar{\bm{r}}) \uu
	\end{align}
	which has no explicit reference to $\bm{H}^{-1}$.  Moreover, note that once the phase of $\uu$ is known one can use the results in Eq. 24 of Ref.~\cite{miatto2020fast} to obtain the net phase of the transformation above.

	\bibliography{bib.bib}
\end{document}